\documentclass[aps,prb,twocolumn,groupedaddress,showpacs]{revtex4}
\usepackage{times}
\usepackage{amsfonts}
\usepackage{amsmath, amsthm, amssymb}
\usepackage{graphicx}
\newcommand{\eg}{\textit{e.g. }}%[syn: f.eks., for example, for instance]
\newcommand{\etal}{\emph{et al.}}

\begin{document}
\title{Superconducting Spintronics with Magnetic Domain Walls}
\author{Jacob Linder$^{1*}$ and Klaus Halterman$^{2*}$}

\affiliation{$^1$Department of Physics, Norwegian University of
Science and Technology, N-7491 Trondheim, Norway}
\affiliation{$^2$Michelson Lab, Physics Division, Naval Air Warfare Center, China Lake, California 93555}
\affiliation{$^*$Both authors contributed equally to this work.}

\begin{abstract}
\textbf{The recent experimental demonstration of spin-polarized supercurrents offer a venue for establishment of a superconducting analogue to conventional spintronics. Whereas domain wall motion in purely magnetic structures is a well-studied topic, it is not clear how domain wall dynamics may influence superconductivity and if some functional property can be harnessed from such a scenario. Here, we demonstrate that domain wall motion in superconducting systems offers a unique way of controlling the quantum state of the superconductor. Considering both the diffusive and ballistic limits, 
we show that moving the domain wall to different locations in a Josephson junction will change the quantum ground state from being in a 0 state to a $\pi$ state. Remarkably, we also show that domain wall motion can be used to turn on and off superconductivity: the position of the domain wall determines the critical temperature $T_c$ and thus if the system is in a resistive state or not, causing even a quantum phase transition between the dissipationless and normal state at $T=0$.  In this way, one achieves dynamical control over the superconducting state within a single sample by utilizing magnetic domain wall motion.}
\end{abstract}
\maketitle

The research fields of spintronics and superconductivity, once disparate, have in recent years been moved closer to one another due to several key discoveries. The unification of these two fields might seem futile at first glance since ferromagnets are spin-polarized whereas the main constituent of a superconductor, the Cooper pair, resides in a spinless singlet state in conventional Bardeen-Cooper-Schrieffer theory \cite{bcs}. Nevertheless, it turns out that the mutual interplay between magnetism and superconductivity opens a rich vista of new physics far beyond the notion that ferromagnetic order has a detrimental influence on superconducting order. Even setting aside for the moment the possibility of intrinsically unconventional spin-triplet superconductors like Sr$_2$RuO$_4$ \cite{maeno} and uranium-based heavy-fermion compounds \cite{saxena, aoki, ucoge} such as UGe$_2$, URhGe, and UCoGe, it has been realized over the last years that proximate structures of ferromagnets and perfectly conventional $s$-wave superconductors can sustain long-ranged and spin-polarized superconducting correlations, even in extreme environments such as half-metallic compounds \cite{keizer}.  

The core principles which make possible such an unlikely synthesis between magnetic and superconducting order are the Pauli principle and symmetry breaking \cite{buzdinrmp, bergeretrmp}. The former dictates that Cooper pairs in superconductors not necessarily are confined to a spinless state, but that a spin-polarized state may arise as long as the overall wavefunction of the pair satisfies fermionic interchange statistics. Such a change in spin-polarization of the Cooper pair can be triggered by considering hybrid structures comprised of ferromagnets and superconductors. Since translational symmetry is explicitly broken at the interface region, the Cooper pair wavefunction becomes a mixture of its original bulk state and a state with new symmetries generated at the interface region \cite{tanaka_prl, eschrig_jltp}. Cooper pairs with electrons that carry the same spin would not be subject to paramagnetic pair-breaking and could in principle propagate for large distances $\sim100$ nm inside the ferromagnet regardless of the strength of the exchange field, limited only by coherence-breaking processes such as inelasticity, spin-flip scattering, and thermal decoherence. 

Precisely such behavior can occur in textured ferromagnets, to be contrasted with monodomain ferromagnets. In fact, such long-ranged and spin-polarized superconducting correlations may arise even from conventional $s$-wave superconductors when a magnetic inhomogeneity of some sort is present \cite{bergeret}. A number of proposals have been put forth in this regard, ranging from multilayered magnetic structures, domain wall
ferromagnets, and interfaces with spin-active scattering \cite{eschrig_prl_03, pajovic, volkov_prb, linder_prb_09, halterman_prl_07}. Experiments have quite recently been able to unambiguously verify the existence of long-ranged supercurrents flowing through textured magnetic structures \cite{keizer,anwar,robinson}. By now, it is then established that the superconducting proximity effect in ferromagnets may become long-ranged and spin-polarized under suitable circumstances. Although this is certainly interesting from a fundamental physics viewpoint, it begs the question: can these spin-polarized superconducting correlations be utilized for some practical purpose? 

Spin-polarized resistive currents are known to play an instrumental role in the field of spintronics. One of their hallmarks in this context is the ability to transfer angular momentum to the magnetic order parameter in a material, an effect known as spin-transfer torque \cite{slon,berger}. One of the most actively pursued research directions in this field is as of today controllable domain wall motion, which may be accomplished via several routes \cite{domwall_review} such as spin-polarized currents, magnons, and external magnetic fields. Now, such domain walls provide the necessary ingredient to generate spin-polarized superconducting correlations as they represent an inhomogeneous magnetization texture. Therefore, one may envision that the generation of spin-polarized supercurrents may be used to obtain a superconducting spin-transfer torque acting on the magnetization of a 
ferromagnet. In particular, the dissipationless nature of the supercurrent flow offers an interesting venue in terms of reduced energy loss and Joule heating, one of the main obstacles, for efficient domain wall motion. 

In this work, we will demonstrate that domain-wall motion in superconducting junctions offer a novel way of exerting control of the quantum ground state of the system. Varying the position of a domain wall with a realistic magnetization profile taking into account magnetic anisotropy and spin stiffness, we demonstrate that the position of the domain wall controls whether the junction is in a 0- or $\pi$-state. In this way, it becomes possible to exert dynamic control over the quantum ground state within a single sample: the motion of the domain wall manipulates the proximity effects
responsible for the oscillatory nature of the superconducting
order parameter in the ferromagnetic (F) region, as well as the
magnetic correlations and destruction of superconductivity
in the superconducting (S) layers. Moreover, we will show that the domain wall dynamics can result in 
an effective superconducting switch,
where the system changes from a resistive state to
a dissipationless one. We compute the critical temperature of a domain wall nanostructure
in the ballistic regime in an entirely self-consistent manner, which is necessary when
it is unknown {\it a priori} what the ground state of the system is.
We find suitable spin switch candidates that transition from a superconducting state 
to a normal one, even at $T=0$, as the domain wall is shifted. These results show that superconducting spintronics via magnetic domain wall motion can be used not only to change the superconducting quantum state, but even turn superconductivity itself on and off.

We first outline the theoretical framework used in our calculations to compute the superconducting quantum ground state. Next, we present analytical and numerical results for 0-$\pi$ transitions both in the ballistic and diffusive regime of transport, including the possibility of switching from a resistive to dissipationless state simply by moving the domain wall. We then give a detailed discussion of our results, including candidate materials for the predicted effects, and experimental feasibility of our proposed setup. Finally, we summarize our findings.\\ 
\text{ }\\

\begin{figure}[t!]
\centering
\resizebox{0.48\textwidth}{!}{
\includegraphics{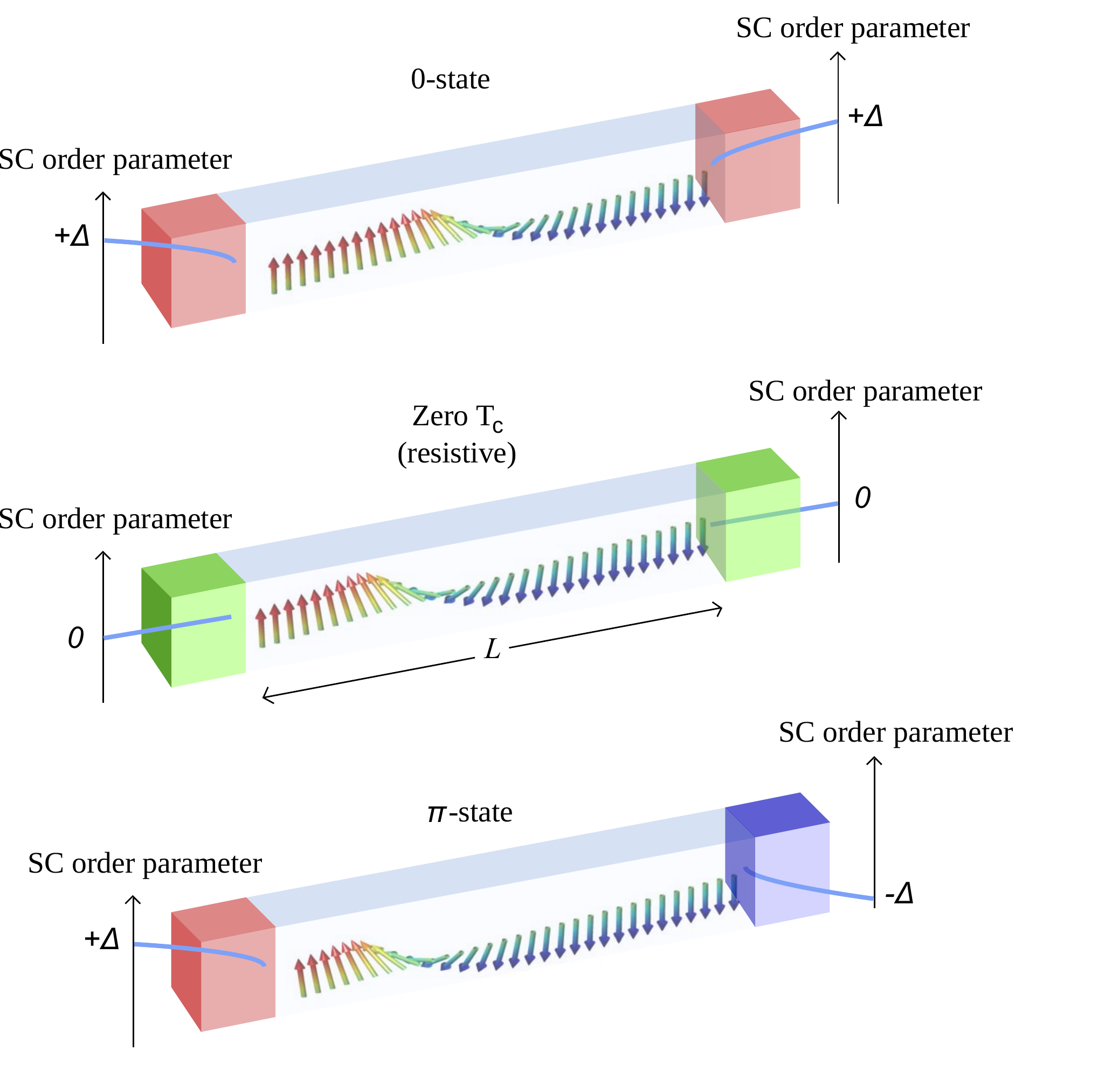}}
\caption{\textbf{Proposed setup}. A magnetic domain wall is present in a ferromagnetic layer of width $L$, separating two conventional $s$-wave superconductors. 
Inducing domain-wall motion to a new position can alter the quantum ground state of the junction, triggering a 0-$\pi$ transition. Moving the domain-wall changes the critical temperature $T_c$ and may even reduce it to zero (middle figure), thus destroying superconductivity. The domain-wall can be moved via an applied current, external field, or spin-wave excitations to specific locations by artificially tailored pinning sites \eg via geometrical notches in the sample. }
\label{fig:model} 
\end{figure}

\noindent\textbf{Domain walls in Josephson junctions}\\
\noindent To model a realistic domain wall, we minimize the free energy functional for an inhomogeneous ferromagnet including exchange stiffness and anisotropy:
\begin{align}
\mathcal{F} = \int \text{d}x[A(\partial_x \mathbf{M})^2/2  - K_\text{easy}M_z^2 + K_\text{hard}M_x^2].
\end{align}
Here, $A$ is the exchange stiffness while $K_\text{easy}$ and $K_\text{hard}$ are the anisotropy energies associated with the easy and hard axes of the magnetization, ${\mathbf M}$, respectively. 
The result \cite{walker} is $\mathbf{M}(x) = [0,\sin\theta(x),\cos\theta(x)]$ where $\theta(x)$ determines the domain wall profile and is given by:
\begin{align}
%\theta(x) = 2\text{atan}\{\text{exp}[(x-X)/\lambda]\}
\theta(x) = 2\arctan\{\text{exp}[(x-X)/\lambda]\}, 
\end{align}
where $\lambda$ is the domain wall width, determined by $A$ and $K_\text{easy}$. We have also introduced the position of the center of the domain wall $X$, which will play an important role in what follows. 
With the magnetization texture in hand, we now insert it into the corresponding equations of motion for the Green's function which in turn enables us to compute the supercurrent in the system. In the diffusive regime, we make use of the quasiclassical Usadel equation \cite{usadel} with the above magnetization profile $\mathbf{M}(x)$:
\begin{eqnarray}
D[\hat{\partial},\hat{G}[\hat{\partial},\hat{G}]]+i[ \varepsilon
\hat{\rho}_{3}+
\text{diag}[\textbf{h}\cdot\underline{\sigma},(\textbf{h}\cdot\underline{\sigma})^{\tau}],\hat{G}]=0.
\end{eqnarray}
Here $D$ is the diffusion constant, $\textbf{h} || \mathbf{M}$ is the exchange field, $\varepsilon$ is the quasiparticle energy, $\hat{\partial}$ is the
derivative  operator,
$\hat{G}$ represents the total
Green's function and $\hat{\rho}_{3}$ and $\underline{\sigma}$ are
$4$$\times$$4$ and $2$$\times$$2$ Pauli matrixes, respectively. The
Usadel equation is supplemented by the Kupriyanov-Lukichev \cite{kl}
boundary conditions at interfaces along the $x$-axis;
\begin{align}
2\zeta\hat{G}\hat{\partial}\hat{G}=[\hat{G}_{\text{BCS}}(\phi),\hat{G}],
\end{align}
in which $\hat{G}_{\text{BCS}}$ is the bulk solution and $\zeta$
controls the interface opacity. For stability in the numerical
computations, we use the so-called Ricatti parametrization \cite{ricatti} of the Green's
function. 
Finally, the supercurrent may be computed according to the formula:
\begin{align}
I_\text{super} = j_0 \int_0^\infty \text{d}\varepsilon\;\text{Tr}\Big\{\hat{\rho}_3 \Big(\check{g}\frac{\partial\check{g}}{\partial x}\Big)^K\Big\}.
\end{align}
where $j_0=-N_0|e|D/16$ is a normalization constant where $N_0$ is the normal-state density of states and $e$ is the electron charge. 
The key observation is that when the Josephson current $I_\text{super}$ changes sign, a 0-$\pi$ transition has taken place. %jl Have now defined the norm. constant j0

We now turn to the ballistic regime to investigate the
transport and thermodynamic properties of
SFS nanojunctions with controllable domain walls.
We utilize
the microscopic Bogoliubov-de Gennes 
(BdG) technique \cite{bogo} which enables us to fully isolate the
superconducting pairing correlations in the system and investigate the precise
behavior of the proximity-induced supercurrent. In
terms of the quasiparticle amplitudes $u_{n \sigma}$ and $v_{n \sigma}$ 
with excitation
energy $\epsilon_n$, and spin $\sigma$,
the BdG equations are compactly written
as,
\begin{align}
&
\begin{pmatrix}  
{\cal H} -h_z&-h_x+ih_y&0&\Delta \\
-h_x-ih_y&{\cal H} +h_z&\Delta&0 \\
0&\Delta^*&-({\cal H} -h_z)&-h_x-ih_y \\
\Delta^*&0&-h_x+ih_y&-({\cal H}+h_z) \\
\end{pmatrix} 
\Psi_n \nonumber \\
%\begin{pmatrix}
%u_{n\uparrow}(x)\\u_{n\downarrow}(x)\\v_{n\uparrow}(x)\\v_{n\downarrow}(x)
%\end{pmatrix}
& \hspace{1cm} =\epsilon_n \Psi_n,
%\begin{pmatrix}
%u_{n\uparrow}(x)\\u_{n\downarrow}(x)\\v_{n\uparrow}(x)\\v_{n\downarrow}(x)
%\end{pmatrix},
\label{bogo}
\end{align}
where
we define the vector 
$\Psi_n\equiv(u_{n\uparrow},u_{n\downarrow},v_{n\uparrow},v_{n\downarrow})^T$. 
The pair potential $\Delta(x)$ 
must be determined self-consistently by solving the BdG
equations together with the condition,
\begin{align} \label{sc}
\Delta(x) = \frac{g}{2}{\sum_{n}}
[u_{n\uparrow}(x)v^*_{n\downarrow} (x)+
u_{n\downarrow}(x)v^*_{n\uparrow} (x)]\tanh\left(\frac{\epsilon_n}{2T}\right),
%\end{widetext}
%\end{subequations}
\end{align}
where the sum is
restricted to those quantum states with positive energies below
an energy cutoff, specified below. %kh2 \omega_D is never mentioned again.
The single particle Hamiltonian ${\cal H}$ is expressed as,
%\begin{equation}
${\cal H}=1/(2m)(-\partial_x^2+k_x^2+k_y^2 )-\mu + U(x)$, %kh minus sign
%\end{equation}
where $\mu$ is the Fermi energy, and $U(x)$ is the spin-independent interface
scattering potential which we take to be of the form $U(x)=U_B [\delta(x+L/2)+\delta(x-L/2)]$, where $L$ is the width of the ferromagnetic region.
The terms $1/(2m)(k_x^2+k_y^2)$ in the Hamiltonian represent the energy of the transverse modes.

To determine the self-consistent ground state of the SFS system, 
one must calculate the free energy, $\cal F$,
given by,
\begin{align}
{\cal F} = - 2T\sum_n \ln \left[2 \cosh\left(\frac{\epsilon_n}{2T}\right)\right]+
%\frac{1}{d_{S1}+d_{S2}}
%\int \limits_{x\in S} dx 
\frac{\langle|{\Delta}(x)|^2\rangle}{g},
\end{align}
where $\langle\ldots\rangle$ denotes spatially averaging
over the entire system, 
and the pair potential is self consistently calculated in Eq.~(\ref{sc}). The supercurrent can be found 
by taking the derivative of the free energy with respect to the 
phase difference $\phi$:
$j_x = 2e ({\partial {\cal F}}/{\partial \phi})$. Note that we have not made any assumption of a weak proximity effect in the above - the results are obtained by solving 
the full proximity effect equations numerically.\\
\text{ }\\

\begin{figure}[t!]
\centering
\resizebox{0.48\textwidth}{!}{
\includegraphics{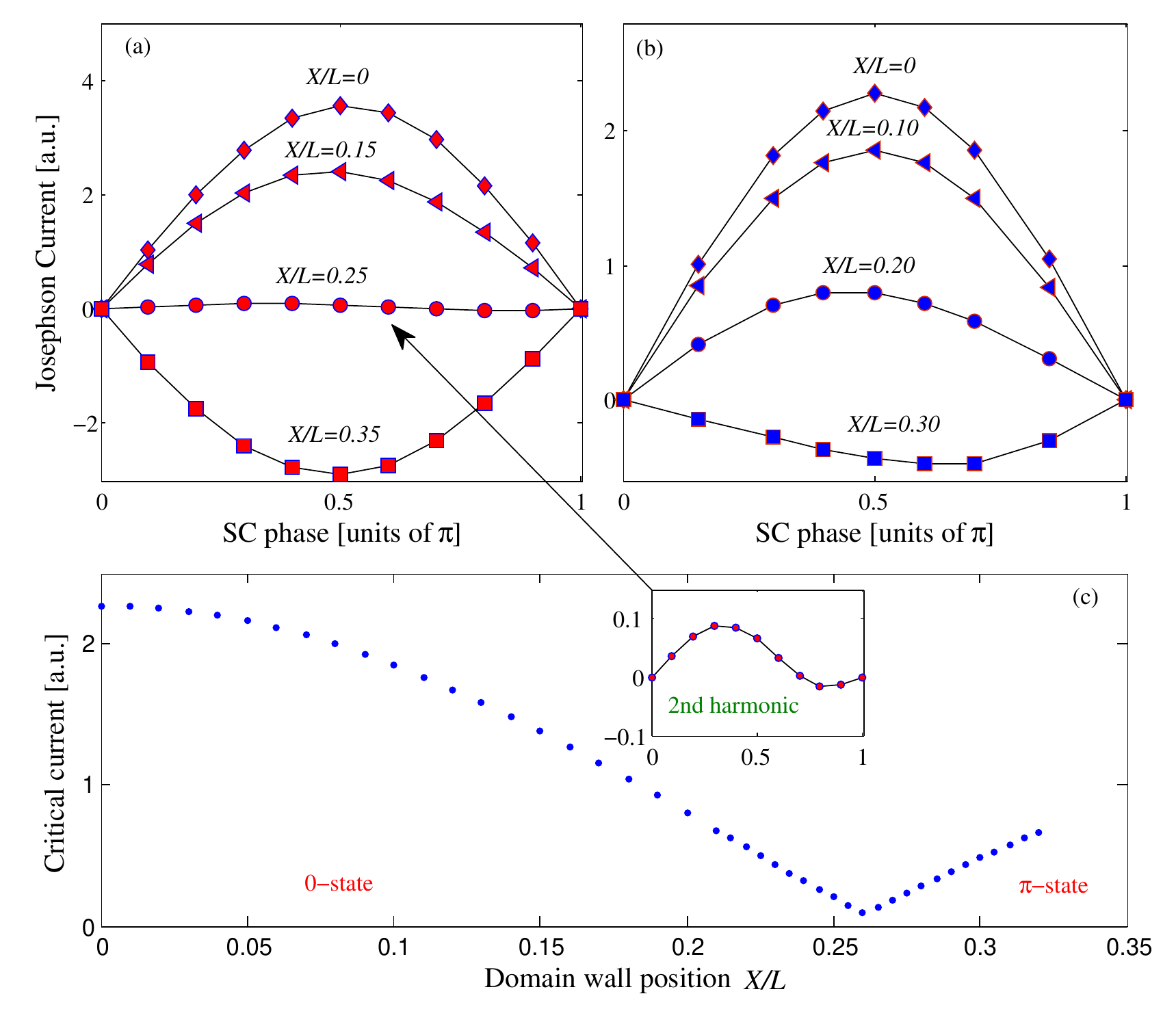}}
\caption{\textbf{Josephson current in the diffusive limit.}. Supercurrent-phase relation for two different parameter sets: (a) $h/\Delta=5, \lambda/L=0.02, L/\xi=1.7$ and (b) $h/\Delta=8, \lambda/L=0.05, L/\xi=1.5$. (c) Critical supercurrent as a function of domain wall position for the same parameter set as in (b). \textit{Inset:} the appearance of a second harmonic in the Josephson relation near the 0-$\pi$ transition. In all cases, we have used an interface parameter $\zeta=4$, corresponding to a weakly transparent interface in terms of tunneling, and a temperature $T/T_c=0.1$.  }
\label{fig:diffusive} 
\end{figure}

\noindent\textbf{Inducing a 0-$\pi$ transition via domain-wall motion}\\
\noindent The underlying physics for this phenomenon may be most easily understood in the limit of a thin domain-wall. In that case, one may think of the ferromagnetic region as an effective bilayer with two ferromagnets in an antiparallel 
configuration. Whether the junction is in a 0- or $\pi$-state is determined by the total phase-shift picked up by an Andreev bound-state carrying the supercurrent through the ferromagnet. This phase-shift depends on the exchange field orientation and the length of the junction. When the ferromagnet 
consists of two regions with antiparallell magnetization, the phase-shift is partially compensated when the bound-state first propagates through one ferromagnet and then the second one with opposite magnetization direction. In fact, when the layers have exactly the same width, the junction is essentially equivalent to an SNS system \cite{hekking}. However, if the layers are allowed to have different thicknesses, the phase-shift picked up by the Andreev bound-state will allow for a $\pi$-state to be formed as long as $h$ and/or $L$ 
are sufficiently large to induce a $\pi$-phase difference as the bound-state makes a full round-trip between the superconductors. We can then qualitatively understand why moving the domain wall will induce 0-$\pi$ transitions: the position of the wall determines the effective phase-shift experienced by the Andreev bound-state as it propagates between the superconductors. When the domain wall is thick, the analogy to a bilayer breaks down since the spin-rotation takes place over a much longer distance. In our approach, we have access to an arbitrary domain wall profile and have verified that the domain wall position still determines whether the junction is in a 
0- or $\pi$-state in the case where the domain wall extends over a large part of the junction.

We start by demonstrating the possibility of having 0-$\pi$ transitions induced by moving the domain wall %kh2 what temperature did you use?
in the diffusive regime.
In Fig. \ref{fig:diffusive}(a) and (b), we have computed the supercurrent-phase relation using two different parameter sets for the sake of showing that this effect does not just occur for special fine-tuned parameters. Fig. \ref{fig:diffusive}(c) illustrates the critical current as a function of the domain wall position $X$ in the ferromagnet. In all plots, the transition is clearly seen: the current-phase relation is inverted whereas the critical current decays towards zero and then rises to finite values. We note that our calculation is done for a scenario where the system has relaxed to equilibrium with the domain wall at position $X$ in the junction, thus corresponding to several measurements of the current (yet within one single sample) with the domain wall at rest in different positions. We will later discuss precisely how this may be accomplished experimentally. An alternative measuring scheme would consist of doing measurements on distinct samples with domain walls at pinned, predetermined locations by means of geometrical notches or other sources of pinning potentials in the ferromagnet \cite{pin}.

For the ballistic results,  
in all cases we have assumed a superconducting
correlation length corresponding to $k_F \xi = 100$ and measure all
temperatures in units of $T_{c0}$, the transition temperature of bulk
S material. We consider  $T=0.01$, except when calculating the critical temperature, %kh2
and fix the energy cutoff at $0.04$, in units of $\mu$.
Scattering at the interfaces is characterized by parameter $Z_B \equiv m U_B/k_F$,
which is unity throughout the calculations unless otherwise noted, corresponding to a moderately transparent interface in terms of tunneling. We have
found however that the domain wall position leading to the $0$-$\pi$ transition
is weakly dependent on $Z_B$. This follows from the fact that 
the magnitude of the exchange interaction 
$h$ shifts the energy
spectrum for spin-up and spin-down quasiparticles by an amount
$-h$ and $+h$ respectively. 
Thus, the oscillatory period of the superconducting correlations in F are primarily determined from 
the difference of the spin-up and spin-down wavevectors, %which are independent of
and not the spin-independent interface scattering. %kh2

\begin{figure}[t!]
\centering
\resizebox{0.48\textwidth}{!}{
\includegraphics{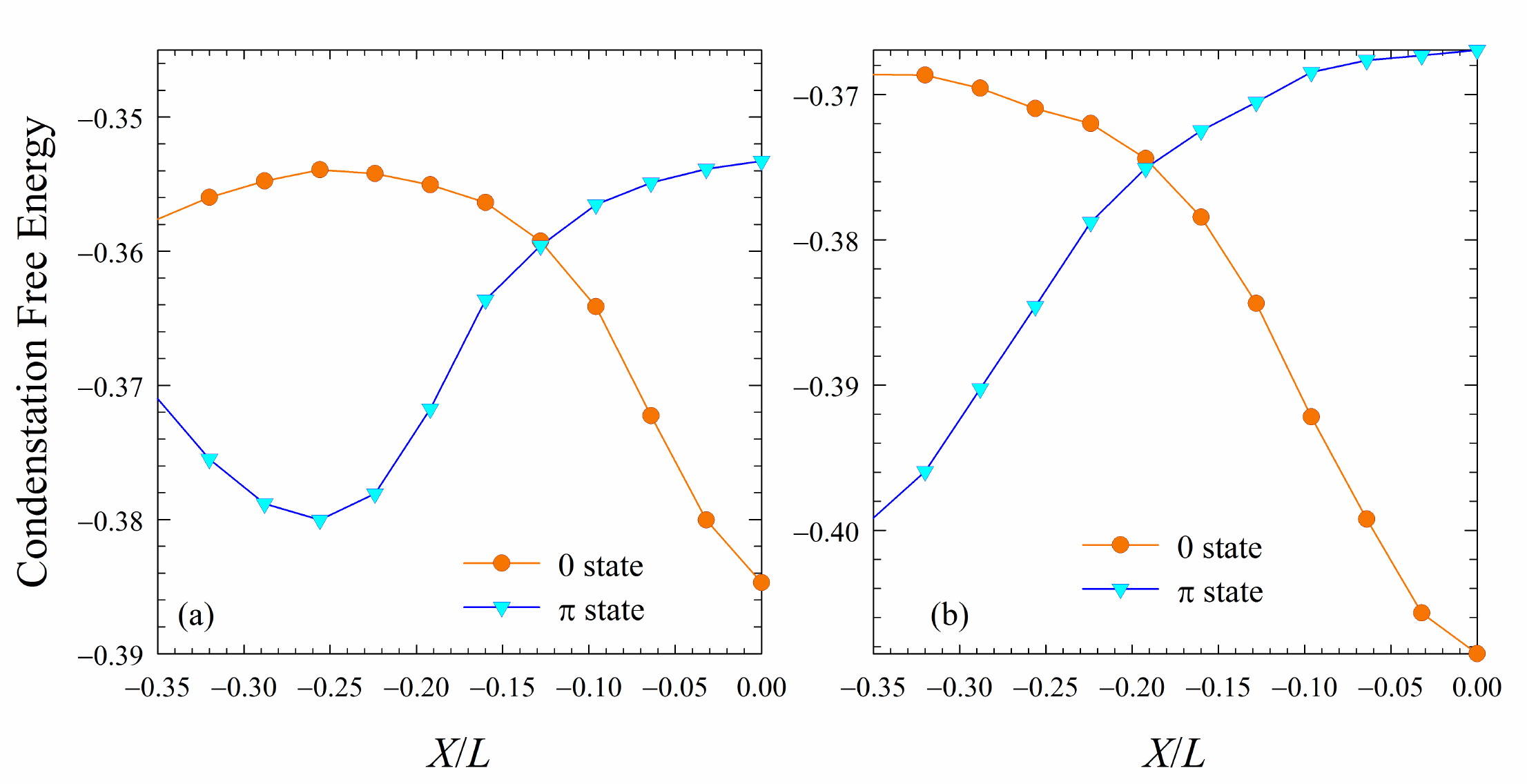}}
\caption{\textbf{Free energy in the ballistic limit}. Control of the quantum state with domain wall motion: (a)
depicts the condensation free energy as a function of domain wall motion for both 
of the
self consistently determined $\pi$ and $0$ states. Here
$\lambda/L=0.02$, 
and $L/\xi=1.5$. (b) corresponds to  $L/\xi=1$ %kh2 is
with all other parameters the same as (a). The center of the junctions corresponds to $X/L=0$
 }
\label{freefig} 
\end{figure}
In the ballistic regime, Fig.~\ref{freefig} 
demonstrates 
the thermodynamics of the $0$-$\pi$ transition,
which follows from the free energy. We
characterize the ground state by 
finding ${\cal F}_S$, the free energy of
the whole system in the self consistent state, and ${\cal F}_N$, 
the normal
state ($\Delta \equiv 0$) free energy. The normalized condensation
free energy is then
$\Delta {\cal F} \equiv ({\cal F}_S-{\cal F}_N)/(2E_0)$,
where $E_0$ is the condensation
energy of bulk S material at $T = 0$. By comparing the condensation energies of
the $0$ and $\pi$ state configurations
as a function of the domain wall position, 
we can therefore  immediately identify 
the ground state of the system. For both Fig.~\ref{freefig}(a) and (b),
we see that when the domain wall is located near the center of the ferromagnet ($X=0$),
the $0$-state is the ground state. However, when the domain wall is moved
closer to the interface, the ground state is the $\pi$-state.

\begin{figure}[t!]
\centering
\resizebox{0.6\textwidth}{!}{
\includegraphics{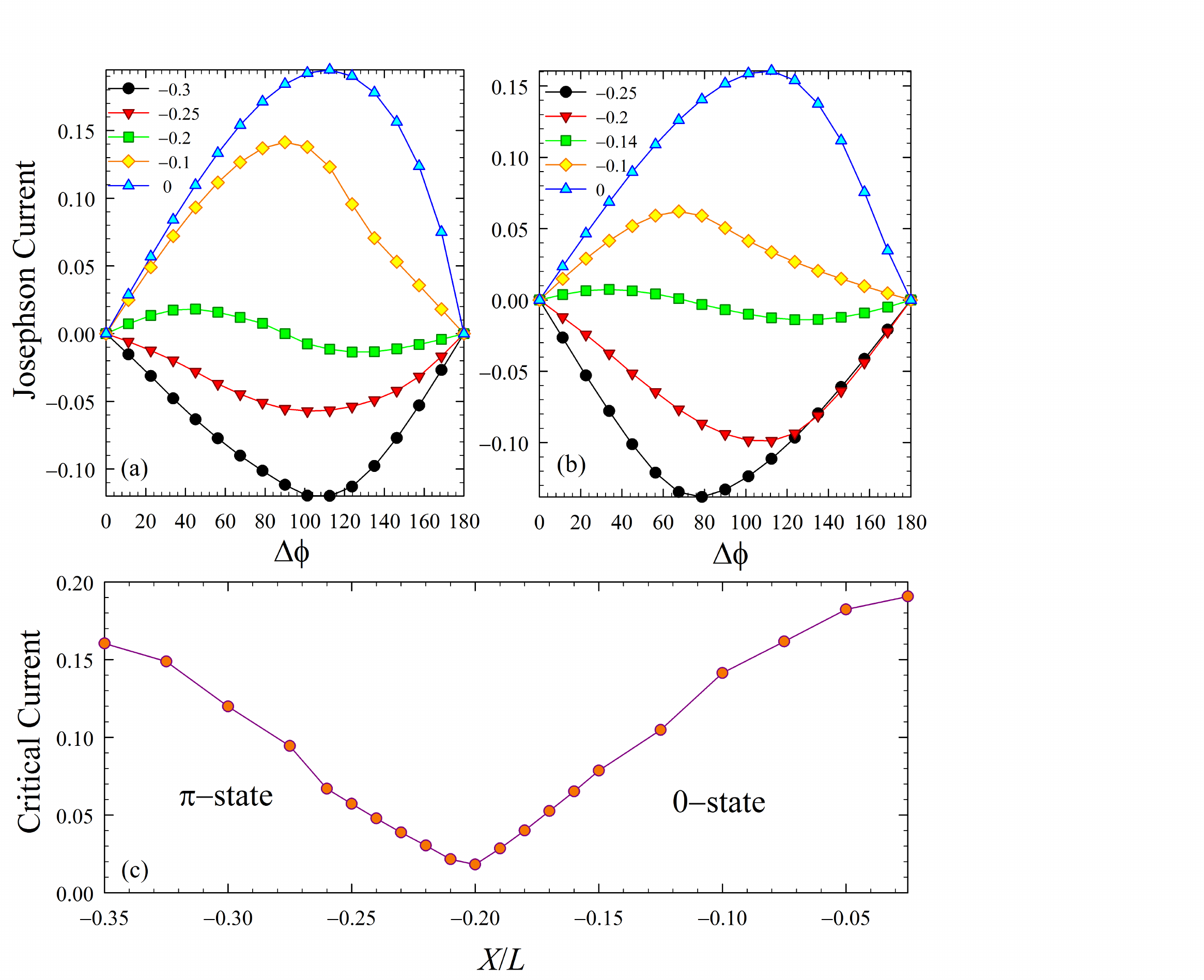}}
\caption{\textbf{Josephson current in the ballistic limit}. Supercurrent-phase relation for two different ferromagnet widths: 
(a) $h/\Delta=5, \lambda/L=0.02, L/\xi=1$ and (b)  $L/\xi=1.5$. 
In both cases there is a clear appearance of a second harmonic in the Josephson 
relation near the 0-$\pi$ transition. 
(c) Critical supercurrent as a function of domain wall position for the same F  thickness 
used in (a). 
 }
\label{jc} 
\end{figure}
Next, in Fig.~\ref{jc} we examine the charge transport and
calculate the Josephson current for the same thicknesses in 
Fig.~\ref{freefig}(a) and (b). The free energy profiles
in Fig.~\ref{freefig}
revealed that the $0$-$\pi$ crossover occurs at $X/L\approx-0.14$ and $X/L\approx-0.2$
for $L=1.5\xi$ and $L=\xi$ respectively.
This is consistent with the supercurrent 
behavior of Fig.~\ref{jc},
where for those domain wall positions, the
current phase relation acquires additional harmonics at
the $0-\pi$ transition.
An experimentally relevant quanitity related to the above results 
is 
the critical current. We therefore show in Fig.~\ref{jc}(c)
the critical current as a function of domain wall position for the cases considered in (a).
This quantity is determined by finding the maximum value of the magnitude of the
Josephson current
over the entire $\Delta \phi$ interval, for each $X$.
The  critical current then has a minimum at
the $0$-$\pi$ transition corresponding to the cusp at  $X/L\approx-0.2$.
Note that  the observed behavior is robust
in the sense that it is found in both the ballistic
and diffusive limit, and for differing thicknesses. 
This should facilitate making contact with experiment.\\
\text{ }\\

\begin{figure}[t!]
\centering
\resizebox{0.45\textwidth}{!}{
\includegraphics{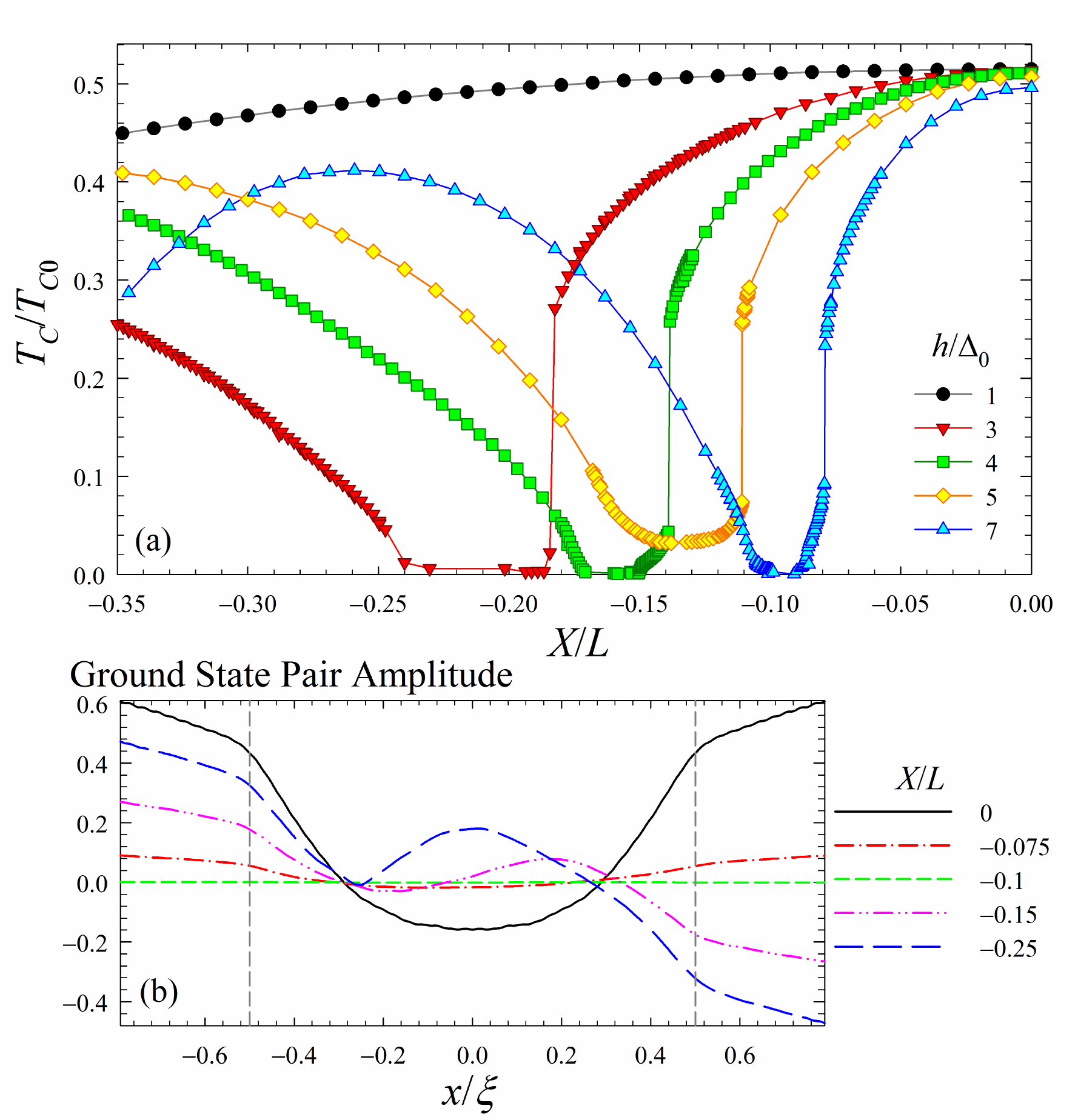}}
\caption{\textbf{Controlling $T_c$ with domain wall motion}. (a) Turning superconductivity on or off: Critical temperature as a function of
domain wall position for several different exchange fields (see legend). 
We assume transparent interfaces and, $\lambda/L=0.02$, $d_S = 0.95 \xi$ %kh2 \xi not \xi_0
and $L/\xi=1$. 
In (b) we show the corresponding Cooper pair
amplitude for the ground states when $T=0$ and for
an exchange field of $h/\Delta_0=7$.
Its spatial dependence reveals the transition from the $\pi$ to $0$ state near the 
minimum of the $T_c$ curve in (a) occurring at $X/L\approx-0.1$, where the system 
has % kh lost its superconductivity and 
transitioned to a normal resistive state.
 }
\label{finaltc} 
\end{figure}

\noindent\textbf{Superconducting on-off switch via domain-wall motion}\\
\noindent We next demonstrate that one can obtain
a superconducting switch controlled by the position of the domain wall.
This effect is revealed
in the experimentally relevant critical temperature, 
which is computed by linearizing the BdG equations (\ref{bogo})
near the transition, yielding an eigenvalue problem that is 
solved using an extension to previous methods. \cite{ilya}
Penetration of the superconducting condensate into the ferromagnet results in the breaking
of  Cooper pairs by the exchange field and leads to
a decrease of the superconducting transition temperature. 

We illustrate in Fig.~\ref{finaltc}(a) the rich variety of switching behavior that can
arise when varying the domain wall position. 
%Multiple values for the magnitude of the exchange field are considered. %kh
Increasing
the exchange field tends to increase the number of $T_c$ oscillations, reflecting the increase
in the period of oscillations in the Cooper pairing amplitude 
that resides in the ferromagnet. 
The critical temperature is typically indifferent 
to $h$ near the center of the junction, where the curves coalesce. 
As the domain wall shifts away from the center, $T_c$ of the system drops abruptly to zero
and the system transitions to a normal resistive state 
in a way  that depends strongly on $h$. This is highly suggestive of a
superconducting switch where superconductivity is turned off or on depending on
the location of the domain wall. 
The application of an external field may also introduce additional interesting reentrance effects \cite{yang}.
It is important to note that
this switching effect is not exclusive to Josephson junctions,
as we have
found that the same effect occurs in a S/F bilayer structure as well. %kh2 

In analogy with the critical current behavior,
the critical temperature contains fingerprints of 
the $0$$-$$\pi$ transition, occurring at around 
the minimum of the $T_c$ curves.
This point is illustrated in Fig.~\ref{finaltc}(b), where we show
the spatial behavior of the Cooper pair amplitude for $h/\Delta_0=7$.
Five differing domain wall positions are considered: two above and two below $X/L\approx-0.1$,
where the system is normal and the pair amplitude vanishes. Clearly, the domain wall
position relative to this transition point dictates whether
the ground state of the system is the $0$ or $\pi$ state.\\
\text{ }\\

\noindent\textbf{Implications for experiments}\\ 
\noindent It is known \cite{domwall_review} that domain wall motion may be induced both via application of a current-induced spin-transfer torque and via external magnetic fields \cite{zhu}. 
Besides these conventional techniques, another possibility was recently unveiled which might be suitable for our purposes. It was demonstrated in Ref.~\onlinecite{yan} 
that domain-wall motion could be obtained via excitation of spin-waves, resulting in a purely magnonic spin-transfer torque. Such spin-waves could be excited via application of a local ac magnetic field $\mathbf{H} = H_0\sin(\omega t)\hat{z}$, giving rise to domain wall motion toward the spin-wave source. Application of such local fields has been successfully implemented experimentally previously \cite{expspinwave}, and might be feasible in our setup as well. Current-induced domain wall motion is also an alternative, 
although 
it might require an additional polarizing ferromagnetic layer in order to achieve an efficient spin-transfer torque. It has been demonstrated that spin-triplet supercurrents can induce magnetization dynamics \cite{linderyoko} and spin-transfer torques \cite{zhao}, and it is thus reasonable to expect that domain wall motion can be induced by a supercurrent spin-transfer torque as well. Once domain wall motion has been induced via \eg one of the above mentioned venues, it is possible to control where the motion terminates, and thus obtain a new ground-state configuration, by artificially tailoring pinning sites which effectively traps the domain wall. This can be accomplished experimentally by \eg making geometrical notches at the desired locations of the ferromagnetic film/wire \cite{pin}. Based on the above discussion, there should then be several alternatives available experimentally in order to move the domain wall in the proposed Josephson junction and thus tune the quantum state of the system to either a 0- or $\pi$-junction and even turn superconductivity on and off. In terms of candidate materials for observation of the predicted effects, one would need two standard $s$-wave superconductors, such as Nb or Al, and a magnetic region supporting a domain wall with a width of order $1-10$ nm. Such domain walls are known to occur in thin magnetic films Pt/Co/AIO$_x$, PtI(Co/Pt)$_n$, and (Co/Ni)$_n$ (see \eg Ref.~\onlinecite{boulle} for a review). Moreover, although we have in our work considered a Bloch-type of domain wall, we do not expect any qualitative change for Neel or head-to-head domain walls, whose textures may be obtained by a rotation in spin space, since the physical principle remains the same. This is advantageous in the sense that generic domain walls, as opposed to a specific type of wall texture, will suffice to experimentally observe the influence on superconductivity predicted here.
\\
\text{ }\\

\noindent\textbf{Conclusion}\\
\noindent We have shown that domain wall motion in superconducting junctions provides a unique way to both tune the quantum ground state between 0$-$ and $\pi-$phases and also turn on and off superconductivity itself. In particular, we find that the domain wall motion may even trigger a quantum phase transition between a resistive and dissipationless state. Our results point towards new ways to merge superconductivity and spintronics in order to achieve functional properties by utilizing domain wall motion. \\
\text{ }\\

\noindent  \textbf{Methods}\\
\begin{scriptsize}
\textbf{Self-consistent calculation in the ballistic limit}. 
It is convenient 
numerically to determine the Josephson
current using the previously calculated 
quasiparticle amplitudes and energies.
Starting with the quantum mechanical expectation value of the 
momentum density, we can can express the current $j_x$ in 
terms of the quasiparticle amplitudes as,
\begin{align}
j_x&=-\frac{ie}{m}\sum_{n,\sigma} \Bigl[u_{n \sigma} \frac{\partial u^{*}_{n \sigma}}{\partial x} f_n+
v_{n \sigma}\frac{\partial v^{*}_{n \sigma}}{\partial x} \left(1-f_n\right)-c.c.\Bigr],
\label{cur}
%jxdown&=jxdown+Ferm*(conjg(udp)*Dudp)+(1.0D0-Ferm)*(vdp*conjg(Dvdp))
\end{align}
where $f_n$ is the Fermi function, $f_n=1/(\exp(\epsilon_n/(2 T))+1)$, and
%   The complete current
%   is then 
%   \begin{align}
%   j^{tot}_{x\sigma}=j_{x\sigma}-j^*_{x\sigma},
%   \end{align}
the $\sigma$ can be either be either spin-up or spin-down ($\uparrow$ or $\downarrow$)
relative to the $z$-quantization axis.
Taking the divergence of the current in Eq.~(\ref{cur}) and using the BdG equations (\ref{bogo}), we find,
\begin{align}
\frac{\partial {j_x}}{\partial x}
&=  
 2i e\Delta({\bm r})\Bigl[ u_{n \uparrow}^*(x) v_{n\downarrow}(x) f_n-u_{n \downarrow}^*(x) v_{n \uparrow}(x)(1-f_n)\Bigr] \nonumber \\
-& 2i e\Delta^*({\bm r}) \Bigl[ u_{n \uparrow}(x) v^*_{n\downarrow}(x) f_n-u_{n\downarrow}(x) v^*_{n \uparrow}(x)(1-f_n)\Bigr].
\end{align}
Thus, when the self-consistency condition is satisfied (Eq.~(\ref{sc})), the right hand side vanishes, and 
the current is conserved. If the self-consistency condition is not strictly satisfied, 
the terms on the right act effectively as sources of current. \\ 
\indent Our numerical procedure for calculating the supercurrent involves
first assuming a piecewise constant form for the pair potential
in each S layer, 
Fourier transforming the real-space BdG equations (Eq.~(\ref{bogo})),
and then diagonalizing the resultant momentum-space matrix.
Once the momentum space wavefunctions and energies are found,
they are transformed back into real-space and
the pair potential is self consistently determined via (\ref{sc}).
The newly calculated $\Delta(x)$ is then inserted back into the BdG equations and the above  
process is repeated.
We generally solve $\Delta(x)$ self consistently 
within about one coherence length
of each side of the domain wall/superconductor interface. 
This leads to the necessary constant current within that region.
Deeper within the S regions, we
take
the self-consistently found $|\Delta(x)|$
and
prescribe a phase difference $\Delta \phi$ across both S banks.
This provides the necessary source of current, and acts as an effective boundary condition.
\end{scriptsize}
\par

\text{ }\\
\noindent  \textbf{Acknowledgments}\\
\begin{scriptsize}
\noindent J.L was supported by the Research
Council of Norway, Grant No. 205591/F20 (FRINAT).  K.H was supported
in part by a grant of HPC resources and the ILIR program sponsored by ONR.
\end{scriptsize}
\par
\text{ }\\
\noindent  \textbf{Author contributions}\\
\begin{scriptsize}
\noindent J.L. and K.H. contributed to the formulation of the problem, theoretical
calculations, and the preparation of the manuscript. 
\end{scriptsize}
\par
\text{ }\\
\noindent  \textbf{Additional information}\\
\begin{scriptsize}
\noindent Correspondence and requests for materials may be addressed to J.L. or K.H.
\end{scriptsize}
\par
\text{ }\\
\noindent  \textbf{Competing financial interests}\\
\begin{scriptsize}
\noindent The authors declare no competing financial interests.
\end{scriptsize}

\end{document}